\documentclass{elsart}

\usepackage{graphics} \usepackage{natbib}
 \journal{social networks}

\usepackage{graphicx}
\usepackage{dcolumn}
\usepackage{bm}
\usepackage{url}

\begin{document}
 \begin{frontmatter}
 \title{How to search a social network}
 \author{Lada Adamic\corauthref{cor1}} \and
 \author{Eytan Adar}
 \corauth[cor1]{ladamic@hpl.hp.com \\
  \url{http://www.hpl.hp.com/research/idl} \\
  tel: (650) 857-5072 \\
  fax: (650)852-8156 }
 \address{HP Labs, 1501 Page Mill Road \\
    Palo Alto, CA 94304}

\begin{abstract}
We address the question of how participants in a small world
experiment are able to find short paths in a social network using
only local information about their immediate contacts. We simulate
such experiments on a network of actual email contacts within an
organization as well as on a student social networking website. On
the email network we find that small world search strategies using
a contact's position in physical space or in an organizational
hierarchy relative to the target can effectively be used to locate
most individuals. However, we find that in the online student
network, where the data is incomplete and hierarchical structures
are not well defined, local search strategies are less effective.
We compare our findings to recent theoretical hypotheses about
underlying social structure that would enable these simple search
strategies to succeed and discuss the implications to social
software design.
\end{abstract}

\begin{keyword}
social networks \sep small world experiment \sep online communities \sep email analysis
\end{keyword}

 \end{frontmatter}
 \pagebreak
 \section{Introduction} Many tasks, ranging from
collaboration within and between organizations, pursuit of
hobbies, or forming romantic relationships, depend on finding the
right people to partner with. Sometimes it is advantageous to seek
an introduction through one's contacts, who could recommend one to
the desired target. Finding such paths through a network of
acquaintances is something people naturally do, for example, when
looking for a job. How people are able to this, while using only
local information about the network, is a question we address by
analyzing real-world social networks. This analysis examines
whether social networks are structured in a way to allow effective
local search. In answering this question we have obtained insights
that may be applicable to new commercial services, such as
LinkedIn, Friendster, and
Spoke\footnote{\begin{flushleft}\url{http://www.linkedin.com/},
\mbox{\url{http://www.friendster.com}},
\mbox{\url{http://www.spokesoftware.com}}\end{flushleft}}, that
have recently sprung up to help people network.

Social networking services gather information on users' social
contacts, construct a large interconnected social network, and
reveal to users how they are connected to others in the
network. The premise of these businesses is that individuals might
be only a few steps removed from a desirable business or social
partner, but not realize it. The services allow their users to get
to know one's friends of friends and hence expand their own social
circle. On a smaller scale, the Club Nexus online community, which
we describe later on in this paper, sought to help students at
Stanford University organize activities and find others with
common interests through their social network.

Although the online social networking trend may be fairly recent,
the observation that any two people in the world are most likely
linked by a short chain of acquaintances, known as the ``small
world'' phenomenon, is not. It has been the focus of much research
over the last forty
years~\citep{killworth78reverse,killworth_pseudomodel79,lundberg75smallworld,milgram67,travers69smallworld}.
In the 1960's and 70's, participants in small world experiments
successfully found paths connecting individuals from Nebraska to
Boston and from Los Angeles to New York . In 2002, 60,000
individuals were able to repeat the experiment using email chains
with an average of $4.1$ links to bridge
continents~\citep{dodds03networks}.

The existence of short paths, which is the essence of the small
world phenomenon, is not particularly surprising in and of itself.
This is due to the exponential growth in people as a function of
distance in the case of random acquaintance networks.  If we take
the average number of acquaintances to be about $c=1000$
(\citet{pool78contacts} estimated in 1978 that the number lies
between 500 and 1,500), one would have $ ~ c^2$ or million
``friends of friends'' and $c^3$ or one billion
``friends-of-friends-of-friends''. This means that it would take
only 2 intermediaries to reach a number of people on the order of
the population of the entire United States. Even with more recent
estimates of personal network sizes of around 300
\citep{mccarty01netsize}, it would take fewer than 3
intermediaries to reach anyone in the country. In fact, the
shortest paths may be contracted even further by the presence of a
few well connected individuals. \citet{newman03ego} has shown that
because one is more likely to know highly connected individuals as
opposed to poorly connected ones, ego centered networks can
achieve sizes significantly larger than, for example, $c^2$ at
distance two.

The above approximations assumed that the network is random. That
is, the overlap in one's friends and one's friends of friends is
negligible. In reality, social networks are far from random. Most
of one's contacts are formed through one's place of residence and
profession, forming tightly knit cliques
\citep{killworth78reverse, killworth_pseudomodel79}. This means
that many of one's friends of friends already belong to the set of
one's own friends. Still, as was shown by
\citet{watts98smallworld} it takes only a few ``random'' links
between people of different professions or location to create
short paths in a social network and make the world ``small''.

Although the existence of short paths is not surprising, it is
another question altogether how people are able to select among
hundreds of acquaintances the correct person to form the next link
in the chain. Participants in small world experiments select a
contact overwhelmingly on geographic proximity and similarity of
profession to the target
\citep{killworth78reverse,bernard_index82}. Recently, mathematical
models have been proposed to explain why these simple strategies
work for forming short paths. The models propose that social
networks need to have a structure such that an individual who is
closer, for example, geographically or professionally, has, with
some probability, a shorter network distance to the target. By
successively choosing individuals whose attributes are "closer" to
the target's attributes, one can rapidly navigate the social
network to form a chain to the target.

The hierarchical network model of \citet{watts2002search} assumes
that individuals belong to groups that are embedded hierarchically
into larger groups. The term "group" refers to any collection of
individuals with which some well-defined set of social
characteristics is associated. For example, an individual might
belong to a research lab, that is in turn part of an academic
department, that is part of a university. The probability that two
individuals have a social tie to one another is proportional to
$e^{-\alpha h}$, where $h$ is the height of their lowest common
branching point in the hierarchy, and $\alpha$ is the decay
parameter. The decay in linking probability means that two people
in the same research laboratory are more likely to know one
another than two people from the same department but different
research labs.

\begin{figure}[tb]
\begin{center}
\includegraphics[scale=0.6]{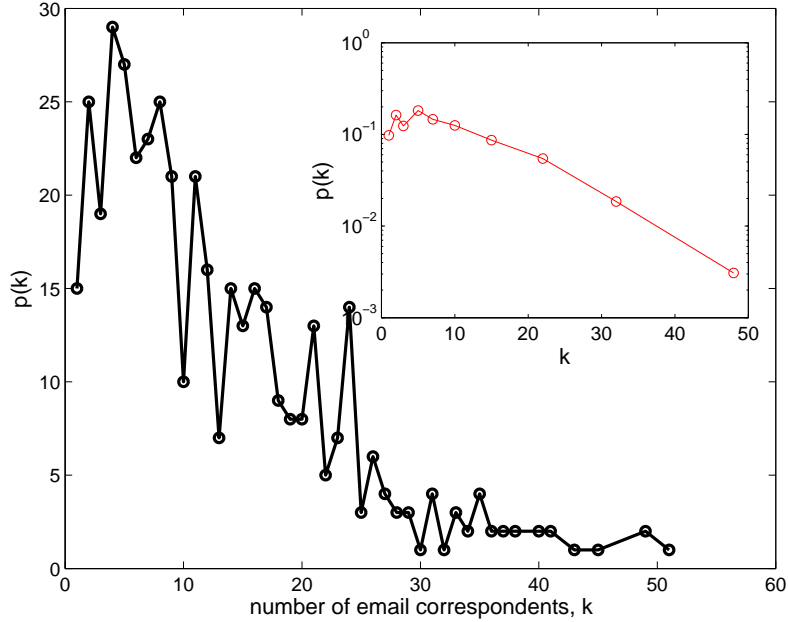}
\end{center}
\caption[Degree distribution in the HP Labs email network ]{Degree
distribution in the HP Labs email network. Two individuals are
linked if they exchanged at least 6 emails in both directions. The
inset shows the same distribution, but on a semilog scale, to
illustrate the exponential tail of the distribution }
\label{swlinkdist}
\end{figure}

Watts et al. applied, on artificial networks, a simple greedy
search strategy that selects the next step in the chain to be the
neighbor of the current node who is closest to the target in an
attribute such as profession or geography. Each node in the chain
has a fixed probability, called the attrition rate, of not passing
the message further. The results of numerical experiments on these
artificially constructed networks showed that for a range of
network parameters, the constructed networks are ``searchable'',
meaning that a minimum fraction of search paths find their target
before attrition terminates them.

\citet{kleinberg2000navigation, kleinberg2001dynamics} posed a related
question: how social networks need to be structured in order for a simple
greedy strategy to find near optimal paths through the network. Unlike the
study of Watts. et al., there is no attrition - all chains run until
completion, but need to scale as the actual shortest path in the network does.
In the case of a small world network, the average shortest path scales as
$\ln(N)$, where $N$ is the number of nodes.

Kleinberg proved that a simple greedy strategy based on geography could
achieve chain lengths bounded by $(\ln N)^2$ if the probability of two
individuals linking is inversely proportional to the square of the distance.
That is, a person would be four times as likely to know someone living a block
away than someone two city blocks away. However, Kleinberg also proved that if
the probabilities of acquaintance do not follow this relationship, nodes would
not be able to use a simple greedy strategy to find the target in
polylogarithmic time. Kleinberg also derived results for individuals belonging
to hierarchically nested groups. If the probability of two people linking to
one another is inversely proportional to the size of the smallest group that
they both belong to, then greedy search can be used to find short paths in
polylogarithmic time.

The models of both Watts et al.\ and Kleinberg show that the
probability of acquaintance needs to be related to the proximity
between individuals' attributes in order for simple search
strategies using only local information to be effective. We will
tie these models to the small world and reverse small world
experiments by simulating on real world networks the strategies
people have reported using and comparing the success of these
strategies with the degree to which the networks conform to a
theoretically searchable structure. Note that we are not examining
here what small world participants actually do, as this has been
the subject of extensive work
\citep{killworth78reverse,killworth_pseudomodel79,lundberg75smallworld,milgram67,travers69smallworld,dodds03networks}.
Rather, we are taking the strategies that participants have
reported using, and are testing, in two experiments, when these
strategies really can be used successfully and why. The first
experiment, described in Section \ref{emailsec} is performed on a
network where social relationships were inferred from email
exchanges and represent a fairly complete picture of the
communication network. The second, described in section
\ref{nexussec} is a search on a network extracted from a social
networking website, containing partial information about the true
network. We also discuss how the differences between networks
constructed through email analysis and the website impact search
performance.

 \section{Search in an Email Network \label{emailsec}}
We first analyzed the email logs at HP Labs to test the
assumptions of the theoretical models regarding the structure of
social networks and the success of simple search strategies. We
took the email network to be fairly representative of the
underlying communication network. A recent study in the UK
\citep{smith03social} found that email was used to communicate
with 80\% of one's social network for the 25-35 age group (60\%
for the 50-60 age group). We expect the percentages to be
significantly higher at HP Labs where email is fairly
indispensable. Although individuals have many ways of
communicating, including face-to-face, over the phone and through
instant messaging, it is unlikely that they communicate without
exchanging any email at all. Email is used to forward information
such as documents, URLs, and other people's email messages, as
well as to schedule face-to-face and phone meetings. Sometimes,
email will be used in tandem with another communication medium
such as voicemail, for example, when an important email is
followed up by a voicemail to make sure that the email is read
\citep{tyler03rhythms}.

We derived a social network from the email logs by defining a
social contact to be someone with whom an individual had exchanged
at least 6 emails both ways over the period of approximately 3
months. Mass emails that are sent to more than 10 individuals at
once were removed. By introducing these thresholds, we sought to
minimize the likelihood of including one sided communication, such
as general announcements, or a very brief email exchange, where
individuals do not get to know one another. The relatively low
threshold of 6 emails still captured weak ties between people in
different departments with little overlap in their social
contacts. Weak links have been shown to play a role job searches
and information diffusion \citep{granovetter73ties}. What we will
show is that these weak ties also play an important role in small
world search. In balance with strong ties that exist within
departments and close physical proximity, they provide a social
network structure favorable to search.

Imposing a communication threshold yielded a network of 430
individuals with a median number of 10 acquaintances and a mean of
12.9. The degree distribution, shown in Figure \ref{swlinkdist},
is highly skewed with an exponential tail. The shape of the
distribution matches that of the estimated network sizes from
scale-up and summation methods of \citet{mccarty01netsize}. The
smaller average number of contacts reflects the restricted setting
of a small organization. The resulting network, consisting of
regular email patterns between HP Labs employees, had 3.1 links
separating any two individuals on average, and a median of 3.

In the simulated search experiments, we considered three different
properties of the nodes: degree, position in the organizational
hierarchy, and physical location. In this simple algorithm, each
individual can use knowledge only of their own email contacts, but
not their contacts' contacts, to forward the message. In order to
avoid passing the message to the same person more than once, the
participants append their names to the message as they receive it,
just as was done in the original experiment by Milgram. We tested
three corresponding strategies, at each step passing the message
to the contact who is
\begin{itemize}
\item best connected

\item closest to the target in the organizational hierarchy

\item located in closest physical proximity to the target

\end{itemize}

The first strategy is a high-degree seeking strategy and selects
the individual who is more likely to know the target by virtue of
the fact that he/she knows so many people. It has been
shown~\citep{adamic01plsearch}, that high degree seeking
strategies are effective in networks with a power-law degree
distribution with an exponent $\gamma$ close to 2. In a power-law
network, the probability of having $k$ contacts is $p(k) \sim
k^{-\gamma}$. This is precisely the degree distribution of the
unfiltered HP Labs email network, where all communication,
including unidirectional and infrequent correspondence, is taken
into account~\citep{wu03flow}. The power-law distribution in the
raw network arises because there are many external senders
emailing just a few individuals inside the organization, and there
are also some individuals inside the organization sending out
announcements to many people and hence having a very high degree.

Once we limit ourselves to emails within HP Labs and impose a
threshold to identify only reciprocal and repeated social
contacts, fewer individuals have a high degree. As we showed
above, the filtered network does not have a power-law degree
distribution, but rather an exponential tail, similar to a Poisson
distribution. \citet{adamic01plsearch} showed that a search
strategy attempting to use high degree nodes in a Poisson network
performs poorly.

Simulation confirmed that the high degree seeking search strategy
was unsuitable for the filtered HP email network. The median
number of steps required to find a randomly chosen target from a
random starting point was 16, compared to the three steps in the
average shortest path. Even worse, the average number of steps was
43. This discrepancy between the mean and median is a reflection
of the skewness of the distribution: the high degree individuals
and their contacts are easy to find, but others who do not have
many links and do not have high degree neighbors are difficult to
locate using this strategy. The unsuitability of the high degree
degree strategy is also intuited by participants in small world
experiments. \citet{bernard_index82} found that contacts were
chosen because they ``knew a lot of people" only 7 percent of the
time. Similarly, \citet{dodds03networks} found that individuals in
successful chains were far less likely than those in incomplete
chains to choose recipients based on their degree (1.6 versus
8.2\%).

\begin{figure}[tbp]
\begin{center}
\includegraphics[scale=0.70]{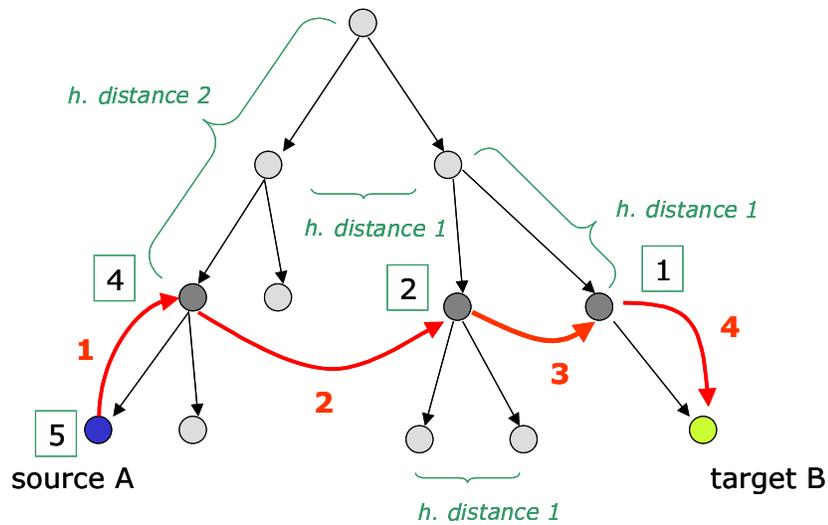}
\end{center}
\caption[Example illustrating a search path using information
about the target's position in the organizational hierarchy
]{Example illustrating a search path using information about the
target's position in the organizational hierarchy to direct a
message. Numbers in the square give the h-distance from the
target. } \label{hierarchyexample}
\end{figure}

\begin{figure}[tb]
\begin{center}
\includegraphics[scale=0.5]{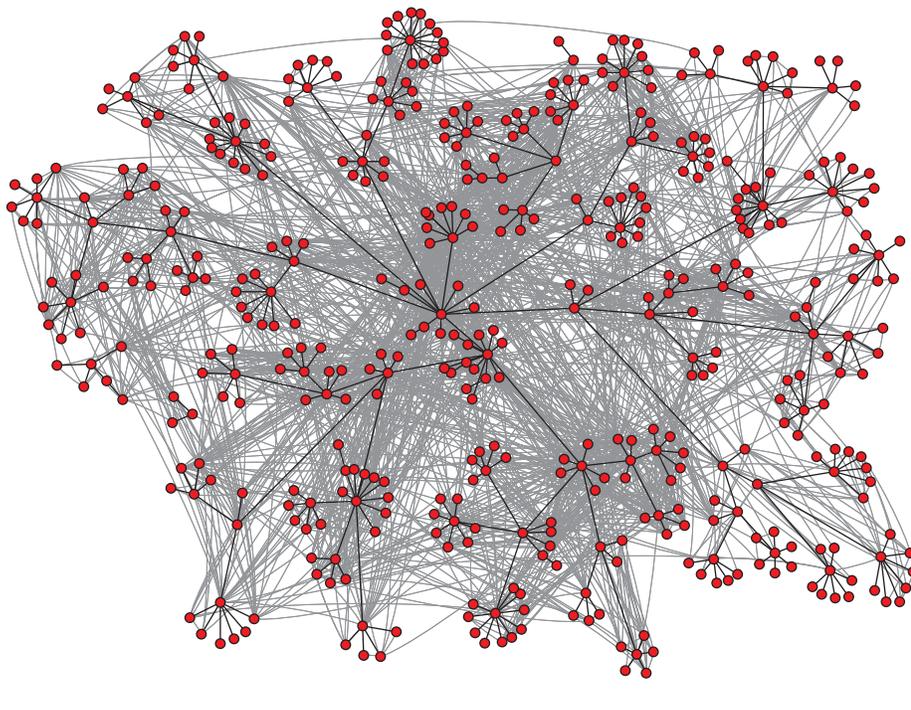}
\end{center}
\caption[Email communications within HP Labs mapped onto the
organizational hierarchy]{HP Labs' email communication (light grey
lines) mapped onto the organizational hierarchy (black lines).
Note that communication tends to ``cling'' to the formal
organizational chart.} \label{hierarchyemail}
\end{figure}

\begin{figure}[tb] \begin{center}
\includegraphics[scale=0.6]{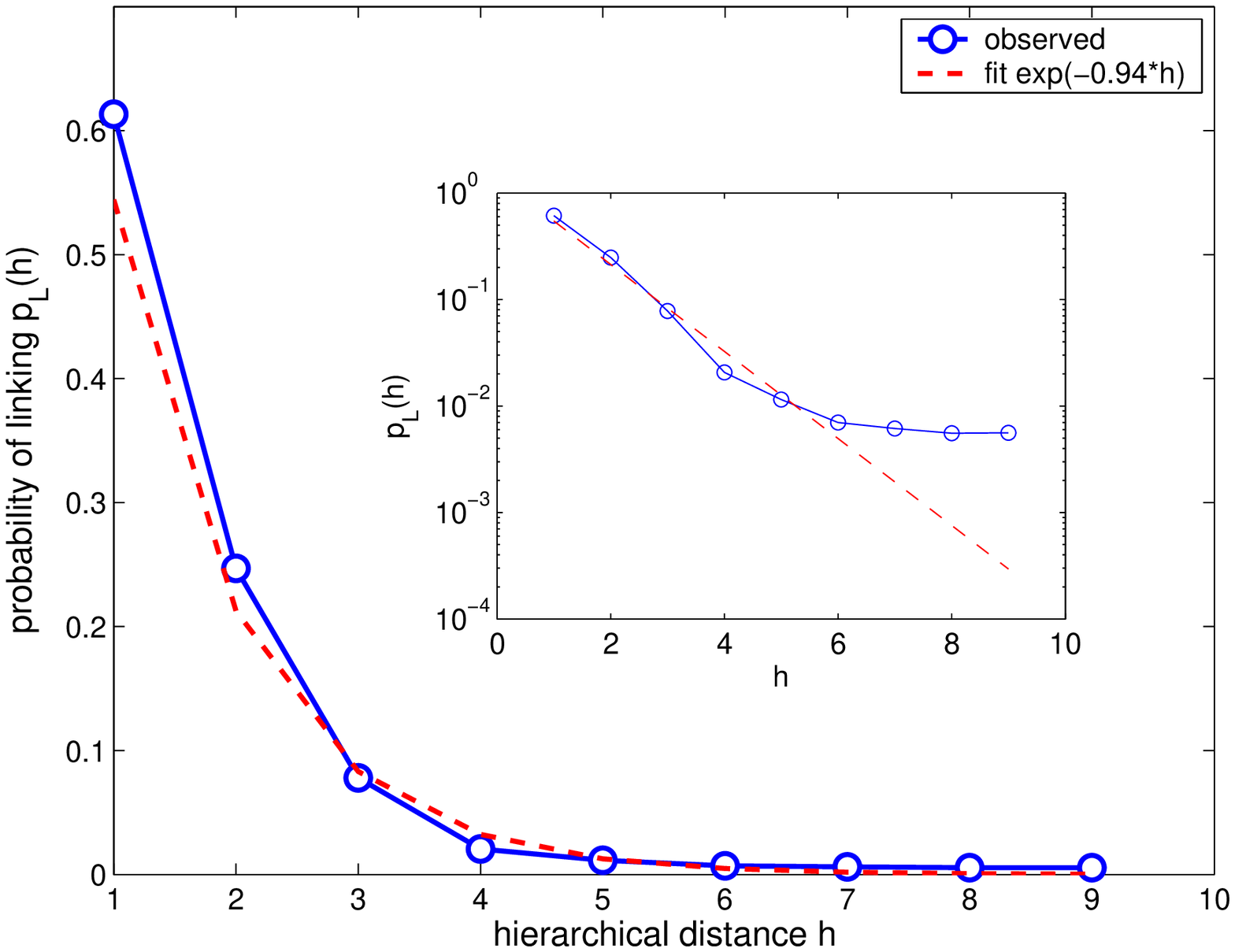} \end{center}
\caption[Probability of linking as a function of the separation in
the organizational hierarchy]{Probability of linking as a function
of the separation in the organizational hierarchy. The exponential
parameter $\alpha = 0.94$, is in the searchable range of the Watts
model~\citep{watts2002search} \label{hierarchylinked}}
\end{figure}

The second strategy consists of passing the message to the contact
closest to the target in the organizational hierarchy. In our
simulation individuals are allowed full knowledge of the
organizational hierarchy (in actuality, employees can reference an
online organizational chart). However, the communication network
that they are trying to navigate is hidden to them beyond their
immediate contacts. Figure \ref{hierarchyexample} illustrates such
a search, labeling nodes by their hierarchical distance
(h-distance) from the target. At each step in the chain the
message is passed to someone closer in the organizational
hierarchy to the target. Note that the message does not need to
travel all the way to the top of the organizational hierarchy and
instead takes advantage of a shortcut created by individuals in
different research groups communicating with one another. The
$h$-distance, used to navigate the network, is computed as
follows: individuals have $h$-distance one to their manager and to
everyone they share a manager with. Distances are then recursively
assigned, so that each individual has $h$-distance 2 to their
first neighbor's neighbors, and $h$-distance 3 to their second
neighbor's neighbors, etc.

The search strategy relies on the observation, illustrated in Figures
\ref{hierarchyemail} and \ref{hierarchylinked}, that individuals closer
together in the organizational hierarchy are more likely to email one another.

\begin{figure}[tb] \begin{center}
\includegraphics[scale=0.6]{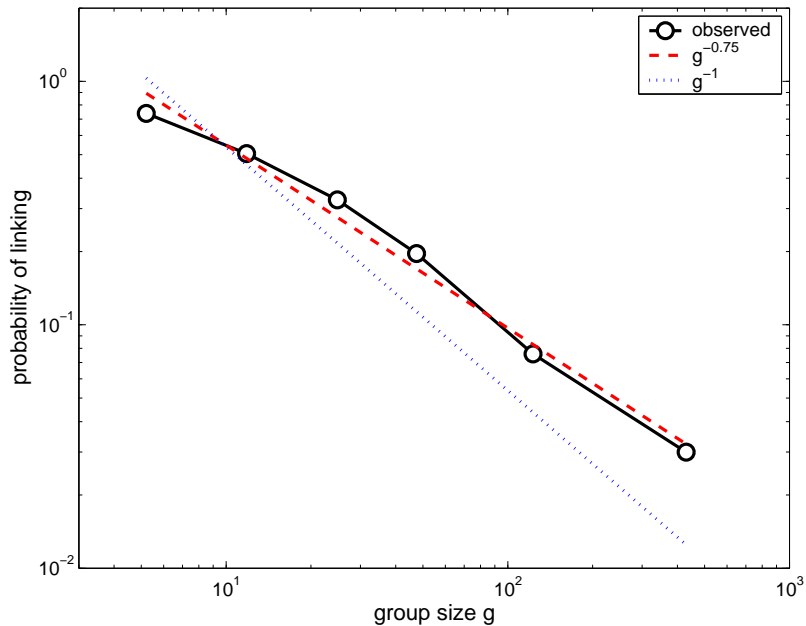} \end{center}
\caption[Probability of two individuals corresponding by email as
a function of the size of the smallest organizational unit they
both belong to.]{Probability of two individuals corresponding by
email as a function of the size of the smallest organizational
unit they both belong to. The optimum relationship derived
in~\citep{kleinberg2001dynamics} is $p \sim g^{-1}$, $g$ being the
group size. The observed relationship is $p \sim g^{-3/4}$. }
\label{groupsizeandlink} \end{figure}

The relationship found between separation in the hierarchy and probability of
correspondence, shown in Figure \ref{hierarchylinked}, is well within the
searchable regime identified in the model of \citet{watts2002search}. However,
the probability of linking as a function of the size $g$ of the smallest
organizational group that both individuals belong to, shown in Figure
\ref{groupsizeandlink}, is a slightly different ($p \sim g^{-3/4}$) from  the
optimal $g^{-1}$\citep{kleinberg2001dynamics}. This means that far-flung
collaborations occur slightly more often than would be optimal for the
particular task of searching, at the expense of short range contacts. The
tendency for communication to occur across the organization was also revealed
in an analysis utilizing spectroscopy methods on the same email network
\citep{tyler03email}. While collaborations mostly occurred within the same
organizational unit, they also occasionally bridged different parts of the
organization or broke up a single organizational unit into noninteracting
subgroups.

Given the close correspondence between the assumptions of the
models regarding group structure and the email network, we
expected greedy strategies using the organizational hierarchy to
work fairly well. Indeed, this was confirmed in our simulations.
The median number of steps was only 4, close to the median
shortest path of 3. The mean was 5 steps, slightly higher than the
median because of the presence of a four hard to find individuals
who had only a single link. Excluding these 4 individuals as
targets resulted in a mean of 4.5 steps. This result indicates
that not only are people typically easy to find, but nearly
everybody can be found in a reasonable number of steps.

\begin{figure}[tb]
\begin{center}
\includegraphics[scale=0.75]{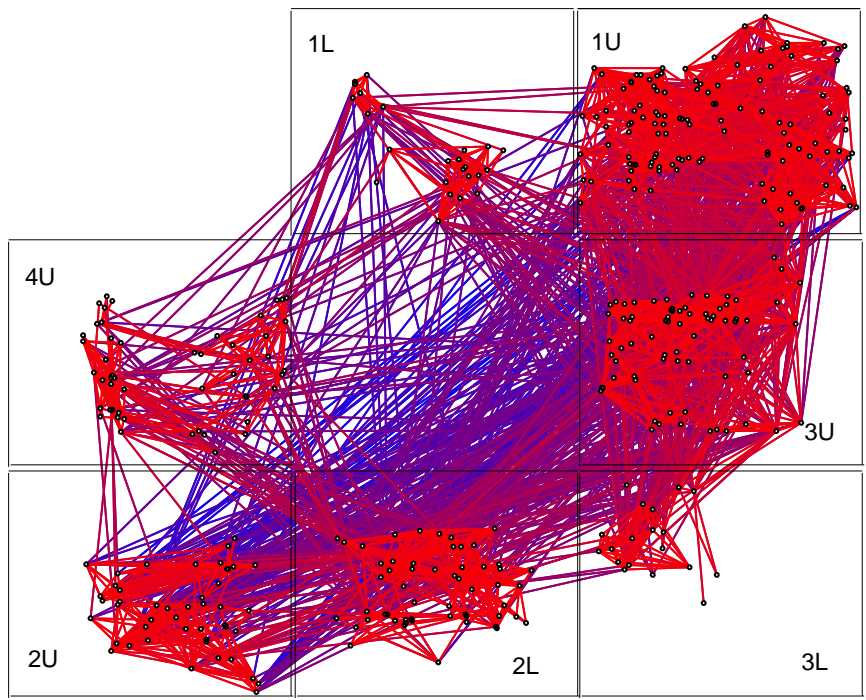}
\end{center}
\caption[Email communications within HP Labs mapped onto
approximate physical location ]{Email communications within HP
Labs mapped onto approximate physical location based on the
nearest post number and building given for each employee. Each box
represents a different floor in a building. The lines are color
coded based on the physical distance between the correspondents:
red for nearby individuals, blue for far away contacts.}
\label{cubiclelayout}
\end{figure}

\begin{figure}[tb]
\begin{center}
\includegraphics[scale=0.6]{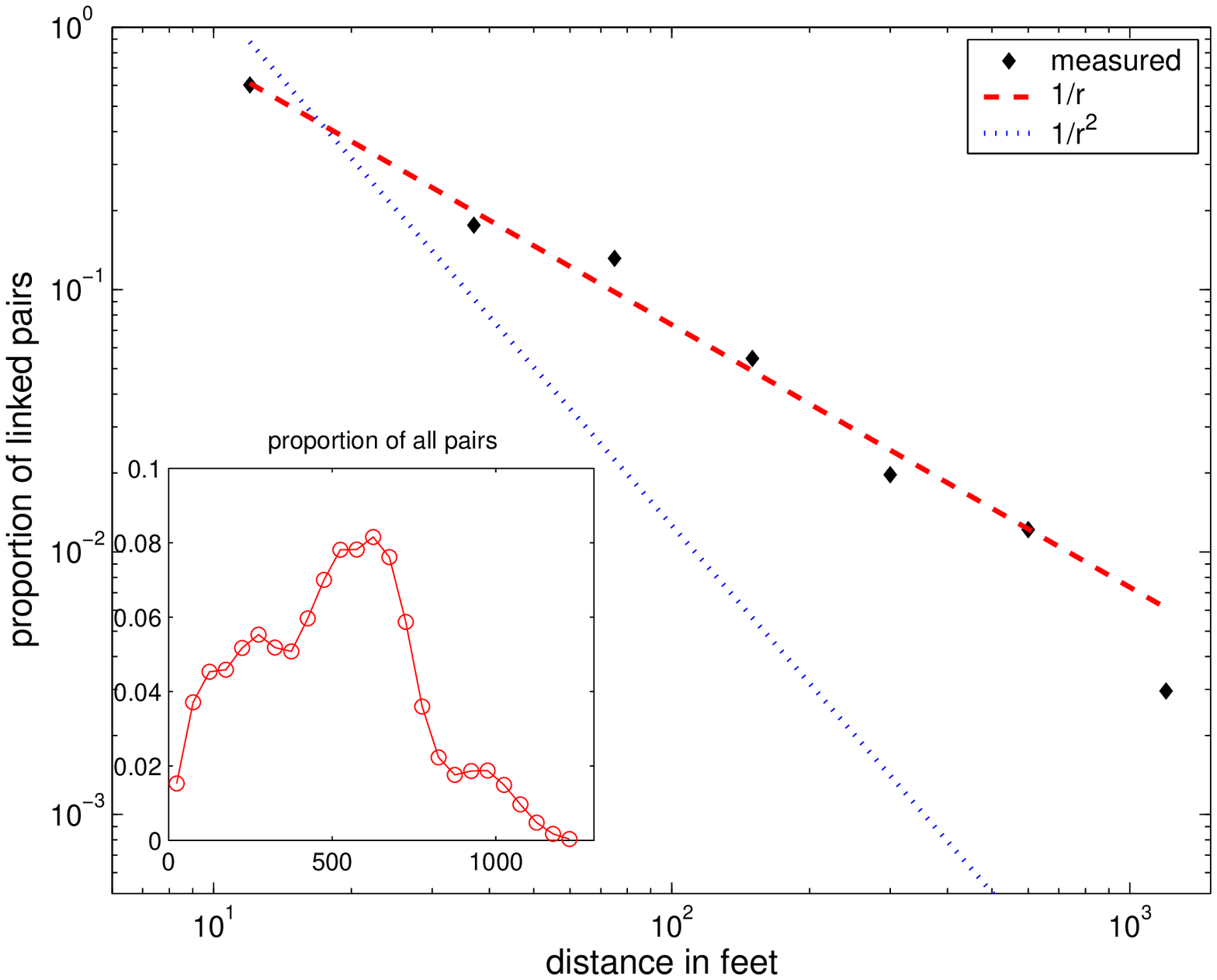}
\end{center}
\caption[Probability of two individuals corresponding by email as
a function of the distance between their cubicles]{Probability of
two individuals corresponding by email as a function of the
distance between their cubicles. The inset shows how many people
in total sit at a given distance from one another.}
\label{cdistandlink}
\end{figure}

The last experiment we performed on the HP email network used the
target's physical location. Individuals' locations are given by their
building, the floor of the building, and the nearest building post
(for example ``H15'') to their cubicle. Figure \ref{cubiclelayout}
shows the email correspondence mapped onto the physical layout of the
buildings.  The general tendency of individuals in close physical
proximity to correspond holds: over 87\% percent of the 4000 email
links are between individuals on the same floor.

We approximate the distance between two cubicles by the ``street''
distance between their posts with building separations and
stairway lengths factored in. The relationship between linking
probability and distance $r$, shown in Figure \ref{cdistandlink},
is $1/r$, different than the optimum $1/r^2$ relationship
Kleinberg derived for two dimensional space. To be fair, the
number of potential contacts at a given distance, shown in the
inset of Figure \ref{cdistandlink}, does not increase as $r^2$ (as
is assumed in Kleinberg's analysis) because of the limiting
geometry of the buildings. Still, the $1/r$ relationship most
likely points to a mild shortage of short-range links. This is the
same inverse relationship found by \citet{Allen77Managing} between
researchers physical separation within an R\&D lab and the
probability that they communicate at least one a week about
technical and scientific matters. We further find that cubicle
distance is correlated, but not completely ($\rho= 0.35$), with
individuals' separation in the organizational hierarchy. A
possible explanation is that the cost of moving individuals from
one cubicle to another when re-organizations occur or new
individuals join, may outweigh the benefit of placing everyone in
the same organizational unit in one location. The availability of
other communication media, such as email, telephone, and instant
messaging, reduces the frequency with which individuals need to
interact face to face, and hence the need to have nearby cubicles.

The shortage of short range links hinders search because one might
get physically quite close to the target, but still need a number
of steps to find an individual who interacts with them.
Correspondingly, our simulations showed that geography could be
used to find most individuals, but was slower, taking a median
number of 6 steps, and a mean of 12. The fact that the strategy
using geography trails behind a strategy using the target's
professional position, is in agreement with Milgram's original
findings. \citet{travers69smallworld} divided completed chains
between those that reached the target through his professional
contacts and those that reached him through his hometown. On
average those that relied on geography took 1.5 steps longer to
reach the target, a difference found to be statistically
significant. The interpretation by Travers and Milgram was the
following: ``Chains which converge on the target principally by
using geographic information reach his hometown or the surrounding
areas readily, but once there often circulate before entering the
target's circle of acquaintances.''

 \begin{table}[tbp]
 \begin{tabular}{l|c|c}
 strategy & median number of steps & mean number of steps \\ \hline
 high degree & 16 & 43.2 \\
 organizational hierarchy & 4 & 5.0 \\
 geography & 6 & 11.7
 \end{tabular}
 \caption{Search times using various strategies. The actual average shortest path is $3.1$.}
 \label{milgramsearchresults}
 \end{table}

Table \ref{milgramsearchresults} summarizes the results of
searches using each of the three strategies. It shows that both
searches using information about the target outperform a search
relying solely on the degree of one's contacts. It also shows the
advantage, consistent with Milgram's original experiment, of using
the target's professional position as opposed to their geographic
location to pass a message through one's contacts.

The simulated experiments on the email network verify the models
proposed by \citet{watts2002search} and
\citet{kleinberg2000navigation} to explain why individuals are
able to successfully complete chains in small world experiments
using only local information. When individuals belong to groups
based on a hierarchy and are more likely to interact with
individuals within the same small group, then one can safely adopt
a greedy strategy - pass the message onto the individual most like
the target, and they will be more likely to know the target or
someone closer to them. At the same time it is important to note
that there are slightly fewer short range contacts both in
physical and hierarchical space, than the optimal proportions
found by \citet{kleinberg2000navigation, kleinberg2001dynamics}.

Our email study is a first step, validating these models on a small scale. It
gives a concrete way of observing how the small world chains can be
constructed. It is quite likely that similar relationships between
acquaintance and proximity (geographical or professional) hold true in
general, and therefore that small world experiments succeed on a grander scale
for the very same reasons.

\section{Searching a network of Friends\label{nexussec}}

The email data we analyzed above provides a fairly complete view
of interpersonal communication within an organization and can be
extracted automatically.  In this section we are interested in
exploring a different kind of network, one constructed from ties
reported by the individuals themselves in the setting of an online
social networking site. We obtained friendship network data from a
community website, Club Nexus \citep{adamic03social}, that allowed
Stanford students to explicitly list their friends. Over 2,000
undergraduate and graduate students joined Club Nexus and listed
their friends as part of the registration process. The online
community provided rich profiles for each of its users, including
their year in school, major, residence, gender, personalities,
hobbies and interests. The richness of the profiles allowed for
detailed social network analysis, including identifying activities
and preferences influencing the formation of friendship
\citep{adamic03social}.

\begin{figure}[tb]
\begin{center}
\includegraphics[scale=0.6]{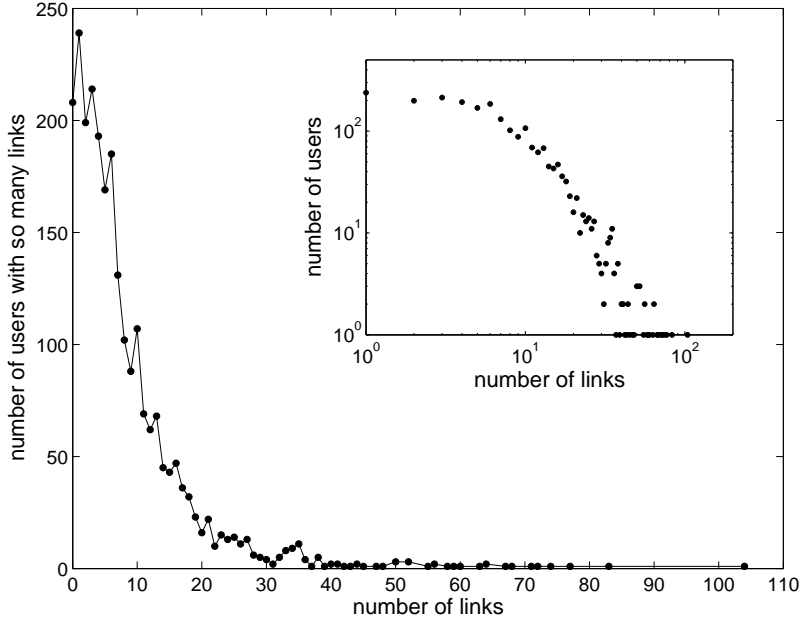}
\end{center}
\caption{Distribution of the number of friends listed by Club
Nexus users. The inset shows the same distribution on a
logarithmic scale.} \label{CNdegdist}
\end{figure}

Club Nexus differed in several respects from the HP Labs email
network, and these differences made it difficult to apply a simple
greedy search strategy to the network. It reflected only a subset
of each persons contacts - those that they would consider friends.
Rather than being able to observe email interactions, as we had
with the HP Labs data set, we had to rely on the users themselves
to list social ties on the site. Upon registering, users were
presented with a ``buddy list" screen where they could add their
friends to their buddy list. The text at the top of the page read:
``Your Buddy List forms the backbone of the Club Nexus system.
From your list of friends, the system will construct your social
network - a required step to enjoy any usage of Club Nexus". Users
could add a buddy either by supplying the buddy's first and last
name and email address, or by searching for them by first or last
name in the database of all Stanford students. Each person added
to the buddy list would then be sent an email asking to
reciprocate the connection by adding the user to their buddy list
and to set up an account on Club Nexus if they had not done so
already. A link that was not reciprocated was deleted from the
system, leaving only bidirectional links that might have been
initiated by either of the two individuals.

The open-ended request for information on one's social network
produced buddy lists varying widely in size. Approximately 209
users specified no friends, and a further 238 listed only 1
friend. On the other end of the distribution, some users had a
'buddy' list containing dozens of friends, resulting in an average
degree of $8.2$. This is in part a reflection of the fact that
some users have more friends than others, but also that some are
more eager to list their friends names, or list more than just
their closest friends, on a website. The full distribution is
shown in Figure \ref{CNdegdist}. Since so many users listed few or
no friends on the website, the Club Nexus social network
represents only a subset of the actual friendships between the
students. The Club Nexus network also differs from the personal
networks typically utilized by subjects in small world
experiments, in that most users' links, in addition to being
restricted to just a subset of their friends, typically do not
include acquaintances.

Given this incomplete network, we tested whether information about
the target could still be used successfully guide a small-world
search. We considered information such as dorm, year, major,
sports, and hobbies of the target in guiding the search, but these
criteria were weak clues compared to information such as position
in the organizational hierarchy and cubicle location that we had
for the HP Labs email network. The clues were weak for several
reasons. The first was due to the lack of acquaintanceship links.
For example, two students living in the same dorm have only about
a $5\%$ probability of being Nexus `buddies', even though one
could assume that the individuals at the very least know each
other by sight. The second difficulty is that closeness in
attributes such as geographical location, does not necessarily
correspond to probability of there being a link, as is required by
Kleinberg's model. In the Club Nexus data, two people living in
different dorms have only a $0.3\%$ chance of being buddies,
regardless of how far apart these residences are. Hence a simple
greedy geographical search, fairly successful in the case of the
HP Labs email network, would not be able to home in on a residence
geographically on the Stanford campus.

\begin{figure}[tb] \begin{center}
 \includegraphics[scale=0.6]{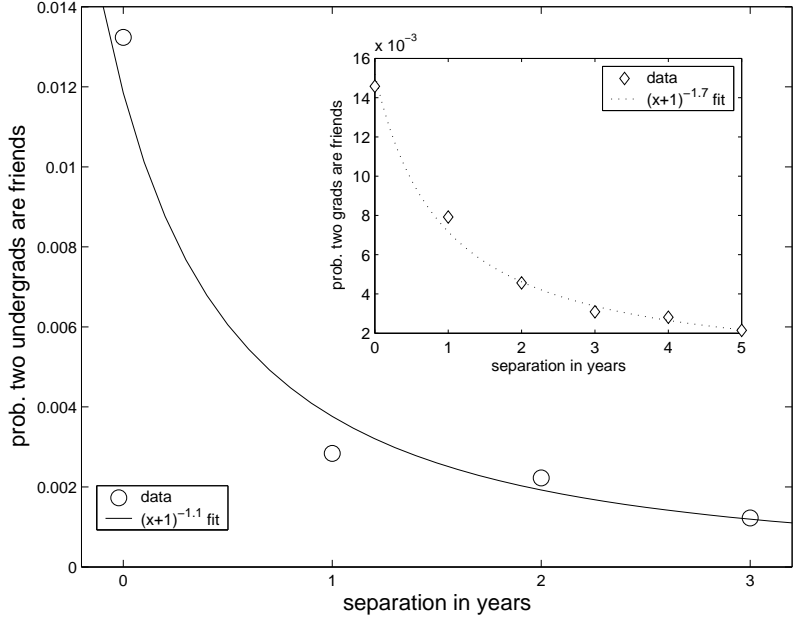} \end{center}
 \caption{Probability of two undergraduates linking as a function of the
difference in their year in school. The inset shows the same for graduate
students.} \label{differenceinyears}
 \end{figure}

Finally, it was difficult to construct a hierarchy for each
attribute as required by the model of \cite{dodds03networks}.
Attributes such as sports or movie genre preferences were
difficult to compare except as exact, binary matches. For example,
it is not clear how similar a swimmer is to a baseball or football
player. For two attributes, department and year in school, we were
able to place attributes in two tier hierarchies. We looked not
only if users belong to the same department but also whether their
departments belong to the same school. For example, two students
might be in the School of Engineering, but one might be a chemical
engineer while the other is a mechanical engineer. We considered
whether students were undergraduates or graduates, followed by
what year they were in. We also took into consideration whether
students were in the same year, or a year apart. Figure
\ref{differenceinyears} shows how likely two individuals are to be
registered as friends on Club Nexus as a function of the number of
years in school separating them. For both undergraduates and
graduates, two people in the same year have approximately a 1\%
chance of being Nexus `buddies', a probability so small as to not
be sufficient on its own to direct the search toward the target.

Our most successful strategy utilizing the user profiles compared 5 attributes
simultaneously for the target and a person one is considering passing the
message to. The possible combinations of attributes were:

 \begin{itemize}
 \item both undergraduate, both graduate, or one of each
 \item same year, a year apart, or more than a year apart
 \item both male, both female, or one of each
 \item same or different residences
 \item same or different major/department and school\end{itemize}

We then calculated the probability that two people know each other (or
have a friend in common) based on all possible combinations of these
variables. The simulation assumes that individuals would be able to
judge, for example, the relative likelihood of someone in the same
year knowing the target, as opposed to someone in a different year,
but in the same dorm. Including additional attributes, such as the
number of shared interests in sports or other activities, did not
improve the search times further.  We experimented with recursively
removing low degree or high degree individuals from the network.  As
can be seen from Table \ref{statstable}, removing low degree nodes
contracts the average shortest path, and makes search shorter as
well. On the other hand, removing very high degree nodes hurts search
by removing connectors that help in completing paths.

We also compared our results to a simple high-degree strategy. For example, if
we keep nodes having 3 or more connections, this leaves us with 1761 users,
with median degree of 6, and average degree of~11. Our strategy requires a
median of 20 steps and a mean of 53 compared to a high degree strategy on the
same network that takes a median number of 39 steps, and 137 on average.

    \begin{table}[tbp]
    \begin{tabular}{c|c|c|c|c}
    link & size of  & average  & median & mean \\
    bracket & connected component & shortest path & \# steps & \# steps \\
    \hline
    $\geq 1$ & 2228 & 4.0 & 28 & 135  \\
    $\geq 2$ & 1992 & 3.8 & 23 & 77  \\
    $\geq 3$ & 1761 & 3.6 & 20 & 54  \\
    $\geq 4$ & 1502 & 3.5 & 18 & 43 \\
    $\geq 5$ & 1204 & 3.3 & 14 & 33 \\
    $\leq 20$ & 1929  & 5.6 & 79 & 231
    \end{tabular}
    \caption{Search times for networks where low degree or high degree nodes are pruned.}
    \label{statstable} \end{table}

Using information about the target outperforms a high-degree
strategy, but the twenty steps required on average are still a far
cry from the six degrees found in real-world small world
experiments. The network is still just barely searchable,
according to the definition provided by \citet{watts2002search}.
Assigning an attrition rate of $r = 0.25$, meaning that at every
step there is a 25\% chance that an individual will not pass the
message on, 5\% of the messages reach their target, in an average
of $4.8$ steps.

What we see is that a simple greedy search is not particularly
effective, especially once one is already confined to a small
geographical space, such as a university campus.  Further, the
contacts that are available to the simulation, most of them being
close friends of the individual, are just a small fraction of the
complete social network. This is in contrast to the observation
made by \citet{granovetter73ties}, that it is the ``weak ties''
that play a disproportionately important role in bridging
different portions of a social network. In the recent small world
study, \citet{dodds03networks} found relationships described as
``casual" and ``not close" to be more frequently used in
successful chains. In the reverse small world experiment performed
by \citet{killworth78reverse}, participants chose on average about
210 different acquaintances to pass the message to various
hypothetical targets. This indicates that individuals must be
selecting from a large arsenal of acquaintances, something the
Club Nexus data set does not provide.

We also speculate that individuals participating in small world
experiments are able to navigate geographical and professional
space by using knowledge not only of a substantial portion of
their immediate social networks, but also by using more
sophisticated strategies that look more than just one step ahead.
\citet{bernard_index82} found that many of the participants in a
reverse small world experiment were thinking two steps ahead, by
selecting individuals who were not associated with the target by
any characteristics of their own but through an association with
an intermediate individual. This is also supported by the work of
\citet{friedkin_horizons83} on horizons of observability that
showed that individuals can be aware of others who are two steps
removed in the network. He found that the probability that
individuals who do not directly interact are aware of each other
increases with the number of paths of length two which connect
them in the network.

This is an important consideration for those constructing social
software websites. Users only allowed to communicate with close,
local neighbors will be at a disadvantage to those with access to
information about their casual acquaintances and second degree
neighbors. It is therefore important for the website to expose
additional information or to suggest contacts based on a global
view.

\section{Conclusion}
In recent years it has become much simpler to harvest large social
networks due to the growing rise in popularity of electronic
communication media such as email, instant messaging and online
communities. By taking advantage of the availability of large
social networks, our small world experiment ties together recent
theoretical results of social network structure that is conducive
to search and the strategies used in previous small world
experiments. The first small world experiments were only able to
trace narrow chains through a social network and to record
individuals strategies. Reverse small world experiments delved
deeper into individuals strategies and isolated individuals'
social networks, but again could not expose the full underlying
social network. Our experiment bridged recent theory on social
network structure and small world experiments by taking entire
observed social networks and applying the strategies that small
world experiment participants report using. We then correlated the
success of the search strategies with the structure of the
underlying social network.

We simulated small world search in two scenarios. The first was
within an organization, where a substantial portion of regular
correspondence is captured through email. The second was an online
community at a university where members of the community volunteer
information about themselves and who their friends are. We
constructed networks from both communities and simulated a
straightforward greedy search on the network - each node passes
the message to a neighbor who is most like the target.

In the case of the email network, the strategies were successful -
messages reached most individuals in a small number of steps, and
using information about the target outperformed simply choosing
the highest degree neighbor. This was due in large part to the
agreement with theoretical predictions by Watts et al.\ and
Kleinberg about optimal linking probabilities relative to
separation in physical space or in the organizational hierarchy.
It may be the case that social ties in general are somewhat less
structured than acquaintanceships within a small organization such
as HP Labs. This could make it more difficult to orient the
search, but is probably compensated by a wider range of
connections available to participants in world-wide small world
experiments.

In the case of the online community, strategies using information
about the target were less successful, but still outperformed a
simplistic high degree search. The limited success of greedy
search is not surprising given that most available dimensions that
could be searched on were binary, with targets either sharing an
attribute or not. The dimensions could not be organized into a
hierarchy that would allow a search to home in on its target.
Geography was also almost a binary variable, since the probability
of students being friends was on average independent on the
separation between their dorms (unless they happened to live in
the same one). Perhaps most importantly, non-participation or
missing data biased the results by hiding connections that could
be used in the search. Individual's social networks (those that
include all of one's friends, family, and acquaintances) are in
general broader than either the HP Labs or Club Nexus personal
networks because they include individuals met through school,
work, neighborhood, and other interests.

For the developers of social software it is important to understand
how different data collection techniques (automated, implicit versus
manual, explicit) impact the resulting social network and how these
networks relate to the real world.  Where the data is incomplete or
reflects non-hierarchical structure, tools that support social search
should assist users by either providing a broader view of their local
community or directly assisting users through a global analysis of the
network data.

 \mbox{}

 \textbf{Acknowledgements}

We would like to thank Orkut Buyukkokten for the Club Nexus data
and Joshua Tyler for helping to map physical distances at HP Labs.
We would also like to thank Kevin Lai, Rajan Lukose and Bernardo
Huberman for their helpful comments and suggestions.

 \bibliographystyle{elsart-harv}
 \bibliography{../../justinitials}
 \end{document}